\documentclass[10pt]{article}

\begin{document}

\title{$4D$ spacetimes embedded in $5D$ light-like Kasner universes}
\author{J. Ponce de Leon\thanks{E-Mail:
jpdel@ltp.upr.clu.edu, jpdel1@hotmail.com}  \\Laboratory of Theoretical Physics, 
Department of Physics\\ 
University of Puerto Rico, P.O. Box 23343,  
San Juan,\\ PR 00931, USA}

\maketitle
\begin{abstract}

We consider spatially homogeneous, anisotropic cosmological models in $5D$ whose line element can be written as 
$dS^2 = {\cal{A}}(u, v)du dv - {\cal{B}}_{i j}(u, v)dx^{i}dx^{j}$, $(i, j = 1, 2, 3)$, where $u$ and $v$ are light-like coordinates. In the case where  ${\cal{B}}_{i j}$ is diagonal,   we construct three families of analytic solutions to the $5D$ vacuum field equations $R_{AB} = 0$ $ (A, B = 0, 1, 2, 3, 4)$. Among them, there is  a family of self-similar homothetic solutions that contains, as a particular case, the  so-called light-like Kasner universes. In this work we provide a detailed study of the different types of $4D$ scenarios that can be embedded in such universes. For the sake of generality of the discussion, and applicability of the results,  in our analysis we consider the two versions of non-compactified $5D$ relativity in vogue, viz.,  braneworld theory and induced matter theory. We find a great variety of cosmological models in $4D$ which are anisotropic versions of the FRW ones. We obtain models on the brane with a non-vanishing cosmological term $\Lambda_{(4)}$, which  inflate {\it \`{a} la} de Sitter without satisfying the classical false-vacuum equation of state.  Using the symmetry of the solutions, we  construct  a class of non-static vacuum solutions on the brane. We also develop {\it static} pancake-like distributions where the matter is concentrated in a thin surface (near $z = 0$), similar to those proposed by Zel'dovich for the  shape of the first collapsed objects in an expanding anisotropic universe. 
The solutions discussed here can be applied in a variety of physical situations.

\end{abstract}

\medskip



\newpage
\section{Introduction}

In recent years there has been an increased interest in theories that envision our spacetime as embedded in a universe with more than four large dimensions. There are several reasons that justify this interest, among them that   
extensions of four-dimensional general relativity to five and more dimensions seem to provide the best route to unification of gravity with the interactions of particle physics \cite{Davidson Owen}-\cite{particle physics}. 
In $5D$ there are  two versions of relativity   where the extra dimension is not assumed to be compactified. These are membrane theory \cite{algebraically} and space-time-matter (or induced matter) theory \cite{general}. They lead to a great variety of models both in the cosmological context and in the description of local self-gravitating objects (see, e.g., \cite{review}, \cite{available}). Most of these models  have been obtained in coordinates where  the metric in $5D$ can be written as\footnote{Notation:  $x^{\mu} = (x^0, x^1, x^2, x^3)$ are the coordinates in $4D$ and $\psi$ is the coordinate along the extra dimension.  We use spacetime signature $(+, -, -, -)$, while $\epsilon = \pm 1$ allows for spacelike or timelike extra dimension, both of which are physically admissible for a detailed discussion see, e.g., \cite{epsilon}.} 
\begin{equation}
\label{General line element in 5D without off-diagonal terms}
dS^2 = g_{\mu\nu}(x^{\rho}, \; \psi)dx^{\mu}dx^{\nu} + \epsilon \Phi^2(x^{\rho}, \; \psi)d\psi^2,
\end{equation}
in such a way that  our $4D$ spacetime can be recovered by going onto a hypersurface $\Sigma_{\psi}: \psi =  \psi_{0} = $ constant, which is orthogonal to the $5D$ unit vector
\begin{equation}
\label{unit vector n}
{\hat{n}}^{A} = \frac{\delta^{A}_{4}}{\sqrt{\epsilon g_{44}}}, \;\;\;n_{A}n^{A} = \epsilon,
\end{equation}
along the extra dimension, and $g_{\mu\nu}$ can be interpreted as the metric of the spacetime.

  In this framework, the effective equations for gravity in $4D$ are obtained from dimensional reduction of the Einstein field equations in $5D$. The reduction is based on Campbell's theorem \cite{Campbell}, \cite{Seahra} and  consists in isolating the $4D$ part of the relevant $5D$ geometric quantities and use them to construct the $4D$ Einstein tensor ${^{(4)}G}_{\alpha \beta}$. The crucial result is that, even in the case where the energy-momentum tensor (EMT) in $5D$ is zero, to an observer  confined to making physical measurements in our ordinary spacetime, and not  aware of the extra dimension,  the spacetime is not empty but contains (effective) matter whose EMT,  ${^{(4)}T}_{\alpha\beta}$, is determined by the Einstein equations in $4D$, namely 
\begin{eqnarray}
\label{4D Einstein with T and K}
{^{(4)}G}_{\alpha\beta} = 8 \pi \;{^{(4)}T}_{\alpha\beta} = 
- \epsilon\left(K_{\alpha\lambda}K^{\lambda}_{\beta} - K_{\lambda}^{\lambda}K_{\alpha\beta}\right) + \frac{\epsilon}{2} g_{\alpha\beta}\left(K_{\lambda\rho}K^{\lambda\rho} - (K^{\lambda}_{\lambda})^2 \right) - \epsilon E_{\alpha\beta}, 
\end{eqnarray}
where $K_{\mu\nu}$ is the extrinsic curvature 
\begin{equation}
\label{extrinsic curvature}
K_{\alpha\beta} = \frac{1}{2}{\cal{L}}_{\hat{n}}g_{\alpha\beta} = \frac{1}{2\Phi}\frac{\partial{g_{\alpha\beta}}}{\partial \psi};  
\end{equation}
$E_{\mu\nu}$ is the projection of the bulk Weyl tensor ${^{(5)}C}_{ABCD}$ orthogonal to ${\hat{n}}^A$, i.e., ``parallel" to spacetime, viz.,
\begin{equation}
\label{Weyl Tensor}
E_{\alpha\beta} = {^{(5)}C}_{\alpha A \beta B}{\hat{n}}^A{\hat{n}}^B 
= - \frac{1}{\Phi}\frac{\partial K_{\alpha\beta}}{\partial \psi} + K_{\alpha\rho}K^{\rho}_{\beta} - \epsilon \frac{\Phi_{\alpha;\beta}}{\Phi},
\end{equation}
and $\Phi_{\alpha} \equiv \partial \Phi/\partial x^{\alpha}$. Before going on, it is worthwhile to emphasize that the above dimensional reduction of the field equations in $5D$ is a standard technique that leads to the same effective matter content in $4D$, i.e., ${^{(4)}T}_{\alpha\beta}$, regardless of whether the line element  (\ref{General line element in 5D without off-diagonal terms}) is interpreted within the context of  brane theory  with ${\bf Z}_2$ symmetry \cite{Shiromizu} or space-time-matter (STM) theory \cite{Wesson and JPdeL}. In this sense these two approaches to   $5D$ relativity are mathematically equivalent.  However, they are different as regards physical interpretation and motivation \cite{physical motivation}. In brane theory there is a singular hypersurface that defines spacetime, and the properties of matter in that hypersurface are, in general,  {\it not identical} to the ones  of induced matter calculated in STM from the  effective EMT defined by (\ref{4D Einstein with T and K}).

In the cosmological realm, nearly all models assume spatial homogeneity and isotropy, which means that the line element in $5D$ is taken to be an extended version of the conventional Friedmann-Roberson-Walker (FRW) metric in $4D$, namely
\begin{equation}
\label{Usual cosmological metric in 5D}
dS^2 = n^2(t, \psi)dt^2 - a^2(t, \psi)\gamma_{i j} dx^{i}dx^{j} + \epsilon \Phi^2(t, \psi) d\psi^2,\;\;\;i, j = 1, 2, 3,
\end{equation}
where $\gamma_{i j}$ is a maximally symmetric $3$-dimensional metric, with curvature index $k = 0, \pm 1$. In these coordinates the full integration of the vacuum Einstein field equations in $5D$ requires the specification of {\it two} additional assumptions. One of them is usually an assumption of geometric nature, e.g., that  $\Phi = 1$, or $n = 1$. The second one is usually an equation of state for the matter quantities in $4D$ \cite{Binetruy}, \cite{JPdeL-isotropicCosm}. 

Observations indicate that on large scales ($\gg 100$ Mpc) the  universe is homogeneous and isotropic and well described by spatially-flat FRW cosmologies. 
However, there is no reason to expect that  such features should  hold  at the early stages of the evolution of the universe. Rather, it is generally accepted that anisotropy could have  played a significant role in the early universe and that it has been fading away in the course of cosmic evolution. In the framework of 4-dimensional spacetime, a prototype for anisotropic vacuum cosmologies is provided by the Kasner metric \cite{Kasner}, which  mimics the behavior of more general solutions near the singularity during some finite periods of time\footnote{Without entering into technical details: extrapolating backwards in time towards the singularity, one finds an infinite number of alternating quasi-periodic Kasner-like epochs with different expansion rates \cite{MTW}-\cite{BKL}.}. Various higher dimensional extensions 
of the vacuum Kasner model have been discussed  in the literature \cite{Kokarev}, \cite{Paul}.

In this paper we consider  spatially homogeneous but anisotropic cosmological models whose metric in  $5D$ has the  form 
\begin{equation}
\label{light-like 5D metrics}
dS^2 = {\cal{A}}(u, v)du dv - {\cal{B}}_{i j}(u, v)dx^{i}dx^{j},
\end{equation} 
 where ${\cal{A}}$ and ${\cal{B}}$ are some functions of the ``light-like" coordinates $u$ and $v$. These metrics  are different from (\ref{Usual cosmological metric in 5D}) in various aspects: (i) they do not contain the time or extra dimension in an explicit way; (ii) the hypersurfaces of constant $u$ or $v$ are three-dimensional instead of $4D$; (ii) {\it{a priori}} it is not clear how to define the $5D$ unit vector ${\hat{n}}^{A}$ along the extra dimension, which in turn is  needed for defining the spacetime sections and for  constructing the appropriate  projected quantities in $4D$. 

Here, for the case where ${\cal{B}}_{i j}$ is diagonal we construct three families of analytic solutions to the $5D$ vacuum field equations $R_{AB} = 0$ $ (A, B = 0, 1, 2, 3, 4)$. The simplest one is a family of self-similar solutions\footnote{In the traditional interpretation of Sedov, Taub and Zel'dovich \cite{Sedov}, self-similarity means that all dimensionless quantities in the theory can be expressed as functions only of a single similarity variable, which is some combination of the independent coordinates. In this way the field equations become a system of ordinary, instead of partial, differential equations.}, which contains as a particular case the so-called light-like Kasner universes. From a physical point of view, self-similar homothetic models are interesting because they   may serve as asymptotic regimes, i.e., near the initial cosmological singularity and at late times, for many homogeneous and inhomogeneous cosmological models \cite{Coley}.  The other two families of solutions are  obtained under the assumption that some of the metric coefficients are separable functions of their arguments.

In view of their potential relevance to the ``similarity hypothesis" \cite{Coley}, in this work we focus our attention to the family of self-similar $5D$ spacetimes mentioned above. The main question under study is what kind of $4D$ scenarios can be embedded in such spacetimes. For the sake of generality of the discussion, and applicability of our results,  in our analysis we consider both versions of non-compactified $5D$ relativity, viz.,   induced-matter and brane theory. Unfortunately, the expressions obtained in $4D$ as projections of the $5D$ solutions are quite complex and cumbersome. Therefore, to obtain manageable mathematical expressions in $4D$, in sections $3$, $4$ and $5$ we simplify the algebra  (but not the physics) by  restricting our discussion to the subset of light-like Kasner solutions. 

Our solutions generalize a number of isotropic cosmological models and give back previous ones in the literature (see e.g., \cite{JPdelW}-\cite{JPdeL-JCAP} and references therein).   Although we are not discussing particular applications here, they  
could be useful in the study of generalizations of Mixmaster or
Belinskii-Khalatnikov-Lifshitz oscillations in theories with a single extra
dimension \cite{mixmaster}, \cite{BKL}, \cite{Halpern}-\cite{Henneaux}. 
They could also be applied
to
studying conjectures about isotropic Big Bang singularities in braneworlds
\cite{Maartens}-\cite{Coley2}.  Certain cosmological models, such as the
cyclic
universe model, also require an understanding of the behavior of
Kasner-like
solutions and are based on brane-type models \cite{Erickson}.

The paper is organized as follows. In section $2$ we derive our self-similar solution (the other two families of solutions are presented in the Appendix) and introduce  a timelike coordinate $t$ and a spacelike coordinate $\psi$ along the extra dimension. This is equivalent to introducing two additional degrees of freedom, which  are expressed in terms of two functions of $t$ and $\psi$. We will see that, as in the familiar FRW picture (\ref{Usual cosmological metric in 5D}),  these two functions can be  related to  the specific choice for    embedding $\Sigma_{\psi}$ in $5D$ and to the physics in $4D$.  In section $3$, within the context of STM we show that  the light-like Kasner solutions generate  a great variety of cosmological models in $4D$, including the de Sitter, Milne and power-law FRW models. 
In section $4$, within the context of the braneworld paradigm we find that they embed $4D$ cosmological models with a non-vanishing cosmological term $\Lambda_{(4)}$, which in principle can be either constant or  time-dependent. In the case of constant $\Lambda_{(4)}$ the $3D$ space exponentially inflates, regardless of the specific embedding. We also show that by virtue of the symmetry $(x, y, z) \leftrightarrow \psi$,  they  generate a class of non-static vacuum solutions on the brane. In section $5$, once again using symmetry properties,  we demonstrate that the Kasner-like metric  (\ref{Lightlike Kasner solution}) can be used to generate static pancake-like distributions of matter in $4D$, similar to those proposed by Zel'dovich for the  shape of the first collapsed objects in an expanding anisotropic universe  \cite{Zel'dovich}. In section $6$ we present a summary of our results.

\section{Cosmological models in $5D$. Light-like coordinates}

\subsection{Solving the field equations. Part I}

In this work we obtain three families of solutions to the field equations $R_{A B} = 0$. However, to facilitate the discussion, in this subsection we present only one of them. Specifically, we present the  family of 
solutions that we will use throughout the paper,   which (as we will see in sections $3$-$5$) may be interpreted or used as $5D$ embeddings for a number of $4D$ universes. The derivation of the other two families of solutions, whose $4D$ interpretation is not discussed here, is deferred  to the Appendix (Solving the field equations. Part II).

To simplify the shape of the field equations let us momentarily denote ${\cal{B}}_{11} = e^{\lambda(u, v)}$,  ${\cal{B}}_{22} = e^{\mu(u, v)}$ and ${\cal{B}}_{33} = e^{\sigma(u, v)}$. From  $R_{xx} = 0$, $R_{yy} = 0$ $R_{zz} = 0$ we obtain  the  equations
\begin{eqnarray}
\label{Rxx, Ryy, Rzz}
4 \lambda_{u v} + \lambda_{u}\left( \sigma_{v} + 2 \lambda_{v} + \mu_{v}\right) + \lambda_{v}\left(\mu_{u} + \sigma_{u}\right) &=& 0, \nonumber \\
4 \mu_{u v} + \mu_{u}\left( \lambda_{v} + 2 \mu_{v} + \sigma_{v}\right) + \mu_{v}\left(\sigma_{u} + \lambda_{u}\right) &=& 0, \nonumber \\
4 \sigma_{u v} + \sigma_{u}\left( \mu_{v} + 2 \sigma_{v} + \lambda_{v}\right) + \sigma_{v}\left(\lambda_{u} + \mu_{u}\right) &=& 0.
\end{eqnarray}
Here the subscripts $u$, $v$ indicate partial derivatives with respect to those arguments. The above equations show cyclic permutation
symmetry, i.e., starting from any of them by means of the transformation $\lambda \rightarrow \mu \rightarrow \sigma \rightarrow \lambda$  we obtain the other two.

$\bullet$ Self-similar solutions: First we solve the field equations under the assumption that the metric (\ref{light-like 5D metrics}) possesses self-similar symmetry. This assumption is motivated by a number of studies  suggesting that self-similar models play a significant role at asymptotic regimes \cite{Coley}. From a mathematical point of view, it means that by a suitable transformation of coordinates all the dimensionless quantities can be put in a form where they are functions only of a single variable (say $\zeta$) \cite{Sedov}. In our particular case, this implies that  $\lambda = \lambda(\zeta)$, $\mu = \mu(\zeta)$, $\sigma = \sigma(\zeta)$, where $\zeta$ is some function of $u$ and $v$, viz.,
\begin{equation}
\label{z}
\zeta = \zeta(u, v). 
\end{equation}
With this assumption the first equation in (\ref{Rxx, Ryy, Rzz}) reduces to
\begin{equation}
\label{equation for lambda in the self-similar solution}
2 \; \frac{\lambda_{\zeta \zeta}}{\lambda_{\zeta}} + \left(\lambda_{\zeta} + \mu_{\zeta} + \sigma_{\zeta}\right) + 2\; \frac{{\zeta}_{u v}}{{\zeta}_{u}{\zeta}_{v}} = 0.
\end{equation}
The assumed symmetry requires $\left({{\zeta}_{u v}}/{{\zeta}_{u}{\zeta}_{v}}\right)$ to be some function of $\zeta$, say $ Z(\zeta) = \left({{\zeta}_{u v}}/{{\zeta}_{u}{\zeta}_{v}}\right)$. Integrating we get
\begin{equation}
\label{lambdaz}
\lambda_{\zeta} = 2 \alpha \; e^{- \left(\lambda + \mu + \sigma\right)/2} e^{- \int{Z(\zeta) d\zeta}} \equiv 2 \alpha \left(\frac{f_{\zeta}}{f}\right),
\end{equation}
where $\alpha$ is an arbitrary constant of integration. Similar equations, with new constants, e.g., $\beta$ and $\gamma$, are obtained for $\mu$ and $\sigma$ by means of a cyclic transformation. What this means is that 
\begin{equation}
\label{lambdaz, muz, sigma z}
\frac{\lambda_{\zeta}}{2 \alpha} = \frac{\mu_{\zeta}}{2 \beta} = \frac{\sigma_{\zeta}}{2 \gamma} = \frac{f_{\zeta}}{f}, 
\end{equation}
which upon integration yields 
\begin{equation}
\label{elambda, emu, esigma}
e^{\lambda}  = C_{1}f^{2 \alpha}, \;\;\;e^{\mu} = C_{2} f^{2 \beta}, \;\;\;e^{\sigma} = C_{3} f^{2 \gamma},
\end{equation}
where $C_{1}$, $C_{2}$, $C_{3}$ are constants of integration. A single differential equation for  $f(\zeta) = f(u, v)$ can be easily obtained by substituting (\ref{elambda, emu, esigma}) into any (\ref{Rxx, Ryy, Rzz}), namely
\begin{equation}
\label{equation for f}
f f_{u v}  + \left(a - 1\right) f_{u} f_{v} = 0. 
\end{equation}

{\it Notation}: Here and henceforth we denote
\begin{equation}
\label{notation}
a \equiv \alpha + \beta + \gamma, \;\;\;b \equiv \alpha^2 + \beta^2 + \gamma^2  - \alpha - \beta - \gamma, \;\;\;c \equiv \alpha \beta + \alpha\gamma + \beta \gamma,
\end{equation}
where $\alpha, \beta, \gamma$ are arbitrary parameters.

A simple integration gives
\begin{equation}
\label{f(u, v)}
f = \left[h(u) + g(v)\right]^{1/a},
\end{equation}
where $h(u)$ and $g(v)$ are arbitrary functions of their arguments. Clearly, in the present case ${\cal{B}}_{i j}$ are power-law type solutions of the  similarity variable $\zeta = \left[h(u) + g(v)\right]$. 

To simplify the discussion, and eliminate spurious degrees of freedom, we now make the  coordinate transformation $h(u) = c_{1} \bar{u}$, $g(v) = c_{2} \bar{v}$, where $c_{1}$ and $c_{2}$ are constants. In these new coordinates
\begin{equation}
{\cal{A}}(u, v)du dv \rightarrow \bar{{\cal{A}}}(\bar{u}, \bar{v})d\bar{u} d\bar{v},
\end{equation}
where $\bar{{\cal{A}}}(\bar{u}, \bar{v}) = \left[{\cal{A}}(u, v)/h_{u}g_{v}\right]$ with $u$ and $v$ expressed in terms of $\bar{u}$, $\bar{v}$. Then relabeling the coordinates (dropping the overbars) the metric  becomes

\begin{equation}
dS^2 = {\cal{A}}(u, v) du d v - C_{1}\left[c_{1} u + c_{2}v\right]^{2\alpha/a}dx^2 - C_{2}\left[c_{1} u + c_{2}v\right]^{2\beta/a}dy^2 - C_{3}\left[c_{1} u + c_{2}v\right]^{2\gamma/a}dz^2.
\end{equation}
 For this metric the field equations $R_{u u} = 0$ and $R_{vv} = 0$ yield the equations
\begin{eqnarray}
\label{Ruu, Rvv}
a^2 \left[c_{1} u + c_{2}v\right]{\cal{A}}_{u} &+& 2 c \; c_{1} {\cal{A}} = 0, \nonumber \\
a^2 \left[c_{1} u + c_{2}v\right]{\cal{A}}_{v} &+& 2 c \; c_{2} {\cal{A}} = 0,
\end{eqnarray}
which have a unique solution given by ${\cal{A}} = C_{0} \left(c_{1} u + c_{2} v\right)^{- 2c/a^2}$, where  $C_{0}$ is a constant of integration. Now, it is easy to verify that $R_{u v} = 0$ is identically satisfied. 
In summary, the final form of the self-similar solution is given by\footnote{The proportionality coefficients $C_{0}$, $C_{1}$, $C_{2}$, $C_{3}$ can be  set equal to unity without any loss of generality.} 
\begin{equation}
\label{new solution}
{\cal{A}} = \left(c_{1} u + c_{2} v\right)^{- 2c/a^2}, \;\;\;{\cal{B}}_{11} =  {\cal{A}}^{- \alpha a/c}, \;\;\;{\cal{B}}_{22} =  {\cal{A}}^{- \beta a/c}, \;\;\;{\cal{B}}_{33} =  {\cal{A}}^{- \gamma a/c}, \;\;\;{\cal{B}}_{i j} = 0, \;\;\;i \neq j.
\end{equation}

We note that this solution admits a homothetic Killing vector in $5D$ for any values of $\alpha$, $\beta$ and $\gamma$, namely, 
\begin{equation}
\label{Lie derivative of the self-similar metric}
{\cal{L}}_{{\zeta}}{g_{A B}} = 2 g_{A B}, \;\;\;\mbox{with}\;\;\;    {\zeta}^C = \left[ {\eta}_{0}  u,\;  {\eta}_{0}  v,\;  (1 - \alpha {\eta}_{0}/a) x,\;  (1 - \beta {\eta}_{0}/a) y,\;  (1 - \gamma {\eta}_{0}/a) z \right],
\end{equation}
where $g_{A B}$ is the metric (\ref{new solution}), ${\cal{L}}_{\zeta}$ denotes the Lie derivative along the $5D$ vector ${\zeta}^C$ and ${\eta}_{0} \equiv a^2/(a^2 - c)$.
In addition,  by setting one of the constants equal to zero, say $c_{2} = 0$, and making the coordinate transformation $u^{- 2c/a^2}du \rightarrow d\bar{u} $, it reduces to 
 \begin{equation}
\label{Lightlike Kasner solution}
dS^2 = d\bar{u} dv -A\bar{u}^{p_{1}}dx^2 - B \bar{u}^{p_{2}}dy^2 - C \bar{u}^{p_{3}}dz^2,
\end{equation}
where $A, B, C$ are constants with the appropriate units, and $p_{1},  p_{2},  p_{3}$ denote 
\begin{equation}
\label{introduction of p1, p2 and p3}
p_{1} = \frac{2 \alpha (\alpha + \beta + \gamma)}{\alpha^2 + \beta^2 + \gamma^2}, \;\;\;p_{2} = \frac{2 \beta (\alpha + \beta + \gamma)}{\alpha^2 + \beta^2 + \gamma^2}, \;\;\;p_{3} = \frac{2 \gamma (\alpha + \beta + \gamma)}{\alpha^2 + \beta^2 + \gamma^2},
\end{equation}
which satisfy the relation $\sum_{i = 1}^{3}\left(p_{i} - 1\right)^2 = 3$
for {\it any} values of $\alpha, \beta $ and $\gamma$. The metric (\ref{Lightlike Kasner solution}) is usually called light-like Kasner solution. In this case the $5D$ homothetic vector is given by ${\zeta}^C = \left[\bar{u},\;  v,\;  (1 - p_{1}/2) x,\;  (1 - p_{2}/2) y,\;  (1 - p_{3}/2) z\right]$.

\subsection{Introducing the timelike and  ``extra" coordinates}

In order to be able to apply the standard dimensional reduction (\ref{4D Einstein with T and K}) to metrics (\ref{light-like 5D metrics}) one has to introduce  coordinates that are adapted to the spacetime sections $\Sigma_{\psi}$.  With this aim we make the coordinate transformation

\begin{equation}
\label{definition of F and V}
u = F(t, \psi), \;\;\;v = V(t, \psi),
\end{equation} 
where $t$ is assumed to be the timelike coordinate; $\psi$ the ``extra" coordinate; $F$ and $V$ are, in principle, arbitrary differentiable functions of their arguments, except from the condition that the Jacobian of the transformation must be  nonzero.

With this transformation we obtain 

\begin{equation}
\label{dudv}
du dv = \left(\dot{F}\dot{V} dt^2 + F' V' d\psi^2\right) + \left(\dot{F} V' + F' \dot{V}\right)dt d\psi,
\end{equation}
where dots and primes denote derivatives with respect to $t$ and $\psi$, respectively. We can choose the coordinates $t, \psi$ in such a way that the $5D$ metric be diagonal. This requires
\begin{equation}
\label{diagonal 5D metric}
V' = - \frac{F' \dot{V}}{\dot{F}}.
\end{equation} 
As a consequence, the line element (\ref{light-like 5D metrics}) becomes
\begin{equation}
\label{light-like metric in t-psi coordinates}
dS^2 = \bar{{\cal{A}}}(t, \psi)\dot{F}\dot{V}dt^2 - {\bar{{\cal{B}}}}_{i j}(t, \psi)dx^{i} dx^{j} - \bar{{\cal{A}}}(t, \psi)\frac{F'^2 \dot{V}}{\dot{F}} d\psi^2, 
\end{equation} 
where $\bar{{\cal{A}}}(t, \psi) \equiv {\cal{A}}(F, V)$ and ${\bar{{\cal{B}}}}_{i j}(t, \psi) \equiv {{\cal{B}}}_{i j}(F, V)$. A couple of points should be noticed here. Firstly, that the physical requirement $g_{00}  > 0$ demands  $\psi$ to be spacelike. Secondly, that  the line element (\ref{light-like metric in t-psi coordinates}) contains  two arbitrary functions,    which are  not present in the original solution (\ref{new solution}). The question is,  why? Is this a mathematical, or gauge,  artifact?  

The answer to this question is that the arbitrary functions in (\ref{light-like metric in t-psi coordinates}) are {\it not}  gauge artifacts. They reflect the  physical reality  that there are many ways of embedding a $4D$ spacetime in $5D$ while satisfying the field equations. If we choose some particular embedding we obtain a differential constraint connecting $V$ and $F$, which allows us to obtain one of them in terms of the other, e.g., $V$ in terms of $F$. Then,  the remaining unknown function, e.g. $F$,   can be  determined from the physics in $4D$.  
As an illustration of the former assertion, let us consider  two  common   embeddings that arise from the choice of the coordinate/reference system.

\paragraph{Gaussian normal coordinate system:}

A popular choice in the literature is to use the five degrees of coordinate freedom to set $g_{4 \mu} = 0$ and $g_{44} = - 1$. This is the so-called `Gaussian normal coordinate system' based on $\Sigma_{\psi}$.   Consequently, in such coordinates $\dot{V} = ({\dot{F}}/{\bar{{\cal{A}}} F'^2})$ and (\ref{diagonal 5D metric}) becomes $V' = - (1/{\bar{{\cal{A}}} F'})$. Now the condition $(\partial \dot{V}/\partial \psi) = (\partial {V'}/\partial t)$ yields
\begin{equation}
\label{obtaining V from F in Gaussian coord.}
a^2\left(c_{1}F + c_{2}V\right) F'' - 2c c_{1} F'^2 = 0.
\end{equation}
If $c_{2} \neq 0$, this equation gives $V(t, \psi)$ 
for  any smooth function $F(t, \psi)$,  and the metric (\ref{light-like metric in t-psi coordinates}) becomes

\begin{equation}
\label{the metric in Gaussian coordinates}
dS^2 = \left(\frac{\dot{F}}{F'}\right)^2 dt^2 - {\bar{{\cal{B}}}}_{i j}(t, \psi)dx^{i} dx^{j} - d\psi^2.
\end{equation}
If $c_{2} = 0$, then (\ref{obtaining V from F in Gaussian coord.}) is an equation for $F$. Integrating  it  we find 
\begin{equation}
\label{F in Gaussian coordinates, c2 = 0}
F = \left[l(t) + \psi h(t)\right]^{a^2/(a + b)},
\end{equation}
where $l(t)$ and $h(t)$ are arbitrary differentiable functions. 
 
\paragraph{Synchronous reference system:}

The choice $g_{00} = 1$ is usual in cosmology: it corresponds to the so-called synchronous reference system where the coordinate $t$ is the proper time at each point. Thus, setting  $\dot{V} = (1/{\bar{{\cal{A}}}}\dot{F})$, the line element  (\ref{light-like metric in t-psi coordinates})  becomes 

\begin{equation}
\label{cosmological homogeneous solution in synchronous coordinates }
dS^2 =  dt^2 - {\bar{{\cal{B}}}}_{i j}(t, \psi)dx^{i} dx^{j} - \left(\frac{F'}{\dot{F}}\right)^2d\psi^2. 
\end{equation}
In these coordinates (\ref{diagonal 5D metric}) reduces to $V' = - (F'/{\bar{{\cal{A}}}}{\dot{F}}^2)$ and $(\partial \dot{V}/\partial \psi) = (\partial {V'}/\partial t)$ yields
\begin{equation}
\label{obtaining V from F in Synchronous coord.}
a^2\left(c_{1}F + c_{2}V\right) \ddot{F} - 2c c_{1} \dot{F}^2 = 0.
\end{equation}
Thus, for $c_{2} = 0$ we get
\begin{equation}
\label{F in the synchronous coordinates}
F = \left[M(\psi) + t N(\psi)\right]^{a^2/(a + b)}, 
\end{equation}
where $M$ and $N$ are arbitrary differentiable functions of $\psi$. For any other $c_{2} \neq 0$, we obtain $V$ from (\ref{obtaining V from F in Synchronous coord.}) after choosing some smooth function $F(t, \psi)$. 

\bigskip

Thus, in principle the function $V$ can be determined if we know $F$. At this point the question arises of whether we can single out   the function $F$ from ``physical" considerations in $4D$. 
Further analysis of the field equations shows that if we assume an equation of state for the  matter in $4D$,  then we obtain an extra differential equation connecting $V$ and $F$, which in addition to (\ref{obtaining V from F in Gaussian coord.}) or (\ref{obtaining V from F in Synchronous coord.}), allows to express  the   solution (\ref{light-like metric in t-psi coordinates}) in terms of $t$ and $\psi$. This is what is required for the $4 + 1$ dimensional reduction of the $5D$ solutions. The general calculations   are straightforward, but the equations are notational cumbersome in both STM and braneworld theory. On the other hand, (\ref{F in Gaussian coordinates, c2 = 0}) and (\ref{F in the synchronous coordinates}) indicate that a great algebraic simplification is attained if $c_{2} = 0$. In fact, in this case  we can re-scale the function $F$, as $F \rightarrow \bar{F}^{a^2/(a + b)}$, after which the solution
(\ref{light-like metric in t-psi coordinates}) with $c_{2} = 0$    reduces to
\begin{equation}
\label{Kasner-like metric in terms of F-bar}
dS^2 = \dot{\bar{F}}\dot{V} dt^2 - A {\bar{F}}^{p_{1}}dx^2 - B {\bar{F}}^{p_{2}}dy^2 - C {\bar{F}}^{p_{3}}dz^2 -       \frac{{\bar{F}}'^2 \dot{V}}{\dot{\bar{F}}} d\psi^2,
\end{equation}
where  $p_{1}, p_{2}, p_{3}$ are the parameters introduced in (\ref{introduction of p1, p2 and p3}). 
In sections $3$, $4$ and $5$ we use this line element, which we call Kasner-like, for illustrating the fact  that physics in $4D$ determines $F$.

\section{Cosmological models  in $4D$. The STM approach}

The aim of this section is to determine $F$ within the context of induced matter theory. To this end we assume   an equation of state for the effective matter quantities. Our results show that the light-like Kasner metrics (\ref{Lightlike Kasner solution}) can be used,  or interpreted,  as   $5$-dimensional embeddings for a number of cosmological models in $4D$ that are spatially anisotropic extensions  of the FRW ones. 

For the Kasner-like metric (\ref{Kasner-like metric in terms of F-bar}), the components of the effective  EMT induced on $\Sigma_{\psi}: \psi = \psi_{0} = $ constant are given by (in what follows we simplify the notation by  omitting  the bar over $F$ in (\ref{Kasner-like metric in terms of F-bar}) and the index $^{(4)}$ in $^{(4)}T_{\mu\nu}$)  

\begin{eqnarray}
\label{EMT}
8 \pi G T_{0}^{0} &=& \frac{a^2 c \dot{F}}{(a + b)^2 \dot{V} F^2}, \nonumber \\
8\pi G T_{1}^{1} &=& \frac{a \left\{(\gamma + \beta)(a + b) F \left[\frac{\ddot{F}}{\dot{F}} - \frac{\ddot{V}}{\dot{V}}\right] - 2 c(\alpha - \beta - \gamma )\dot{F}\right\}}{2(a + b)^2 \dot{V} F^2}, \nonumber \\
8\pi G T_{2}^{2} &=& \frac{a \left\{(\gamma + \alpha)(a + b) F \left[\frac{\ddot{F}}{\dot{F}} - \frac{\ddot{V}}{\dot{V}}\right] - 2 c(\beta - \alpha - \gamma )\dot{F}\right\}}{2(a + b)^2 \dot{V} F^2}, \nonumber \\
8\pi G T_{3}^{3} &=& \frac{a \left\{(\alpha + \beta)(a + b) F \left[\frac{\ddot{F}}{\dot{F}} - \frac{\ddot{V}}{\dot{V}}\right] - 2 c(\gamma - \alpha - \beta )\dot{F}\right\}}{2(a + b)^2 \dot{V} F^2}.
\end{eqnarray}
Certainly the specific shape of the EMT depends on the embedding. However, there are a number of relationships, between the components of the EMT,  which are ``embedding-independent". These are
\begin{eqnarray}
\label{rel between the stresses}
(\gamma - \beta)T_{1}^{1} + (\alpha - \gamma)T_{2}^{2} + (\beta - \alpha)T_{3}^{3} = 0,
\end{eqnarray}
and
\begin{eqnarray}
\label{rel between T0 and the stresses}
(\alpha + \gamma)T_{1}^{1} -  (\beta + \gamma)T_{2}^{2} &=&  (\beta - \alpha)T_{0}^{0},\nonumber \\
(\alpha + \beta)T_{1}^{1} -  (\beta + \gamma)T_{3}^{3} &=&  (\gamma - \alpha)T_{0}^{0},\nonumber \\
(\alpha + \beta)T_{2}^{2} -  (\alpha + \gamma)T_{3}^{3} &=&  (\gamma - \beta)T_{0}^{0}.
\end{eqnarray}

Let us notice some particular cases:  (i) If two of the parameters are equal to each other (axial symmetry), say $\alpha = \beta$,  then $T_{1}^{1} =  T_{2}^{2}$; (ii) If $T_{1}^{1} =  T_{2}^{2}$ but $\alpha \neq \beta$, then $T_{1}^{1} =  T_{2}^{2} = T_{3}^{3} = - T_{0}^{0}$; (iii) If $\alpha = - \beta$, then $T_{3}^{3} = -  T_{0}^{0}$; (iv) In the case of isotropic expansion $(\alpha = \beta = \gamma)$ then $T_{1}^{1} =  T_{2}^{2} = T_{3}^{3}$ (but not necessarily $T_{1}^{1} =  T_{2}^{2} = T_{3}^{3} = - T_{0}^{0}$).

\subsection{Perfect Fluid}

Let us consider the case where the effective EMT behaves like a perfect fluid. From (\ref{EMT}) we find that 
$T_{1}^{1} =  T_{2}^{2} = T_{3}^{3}$ requires
\begin{equation}
\label{perfect fluid case}
(a + b)\left(\frac{\ddot{F}}{\dot{F}} - \frac{\ddot{V}}{\dot{V}}\right) + 4 c \left(\frac{\dot{F}}{F}\right) = 0,
\end{equation}
which implies $\dot{V} \propto \dot{F} F^{4 c/(a + b)}$. Substituting this into (\ref{diagonal 5D metric}) and using $(\partial \dot{V}/\partial \psi) = (\partial {V'}/\partial t)$ we find that $F$ must satisfy the equation\footnote{We note that (\ref{diagonal 5D metric}) remains invariant under the re-scaling $F \rightarrow \bar{F}^{a^2/(a + b)}$.}
\begin{equation}
\label{equation for F, perfect fluid}
(a + b)F \dot{F}' + 4 c \dot{F}F' = 0,
\end{equation}
from which we get
\begin{equation}
\label{F for perfect fluid}
F = \left[f(t) + g(\psi)\right]^{(a + b)/(4 c + a + b)},
\end{equation}
where $f(t)$ and $g(\psi)$ are arbitrary functions of their arguments. The effective energy density $\rho^{(eff)} \equiv T_{0}^{0}$ and pressure $p^{(eff)} \equiv - T_{1}^{1} = - T_{2}^{2} = - T_{3}^{3}$ are given by 
\begin{equation}
\label{density and pressure, perfect fluid}
\rho^{(eff)} = p^{(eff)}, \;\;\;8\pi G \rho^{(eff)} = \frac{c a^2}{(a + b)^2 F^{2a^2/(a + b)}}.
\end{equation}

\subsection{Ultra-relativistic matter and radiation}

It is well-known that in the case of radiation as well as for ultra-relativistic matter (i.e., particles with finite rest mass moving close to the speed of light) the trace of the EMT vanishes identically. From (\ref{EMT}) we find that $T = T_{0}^{0} + T_{1}^{1} + T_{2}^{2} + T_{3}^{3} = 0$ requires
\begin{equation}
\label{ultra-relativistic matter}
(a + b)\left(\frac{\ddot{F}}{\dot{F}} - \frac{\ddot{V}}{\dot{V}}\right) + 2 c \left(\frac{\dot{F}}{F}\right) = 0.
\end{equation}
This equation is the analogue of (\ref{perfect fluid case}). Following the same procedure as above we find $F = \left[f(t) + g(\psi)\right]^{(a + b)/(2 c + a + b)}$. Therefore, the solution for radiation-like matter resembles that of perfect fluid in the sense that the effective stresses $p_{x}^{(eff)} \equiv - T_{1}^{1}$,  $p_{y}^{(eff)} \equiv - T_{2}^{2}$, $p_{z}^{(eff)} \equiv - T_{3}^{3}$ are proportional to the energy density, viz.,
\begin{equation} 
\label{effective stresses, radiation-like solution}
p_{x}^{(eff)} = n_{x}\rho^{(eff)}, \;\;\;p_{y}^{(eff)} = n_{y}\rho^{(eff)}, \;\;\;p_{z}^{(eff)} = n_{z}\rho^{(eff)}, 
\end{equation}
where $n_{x}, n_{y}$ and $n_{z}$ are constants satisfying $n_{x} + n_{y} + n_{z} = 1$. If we average over the three spatial directions, this is equivalent to saying that the equation of state is ${\bar{p}}^{(eff)} = \rho^{(eff)}/3$, where ${\bar{p}}^{(eff)} \equiv - T_{i}^{i}/3$.

\subsection{Barotropic linear equation of state }

For the sake of generality, and in order to keep contact with isotropic FRW cosmologies, let us study the scenario where the effective mater  is barotropic, that is the ratio $\bar{p}^{(eff)}/\rho^{(eff)}$ is constant.
Thus we set 
\begin{equation}
\label{barotropic equation of state}
\bar{p}^{(eff)} = n \rho^{(eff)},\;\;\;n = \mbox{constant},
\end{equation}  
which for $n = 1$ and $n = 1/3$ gets back the above-discussed perfect fluid and radiation-like scenarios, respectively. Substituting into (\ref{EMT}) we obtain an equation similar to (\ref{perfect fluid case}) and (\ref{ultra-relativistic matter}), but  with the  coefficient $(1 + 3n)$ in front of the term $c\dot{F}/F$. Consequently, $\dot{V} \propto \dot{F} F^{[c(1 + 3n)/(a + b)]}$. The condition $(\partial \dot{V}/\partial \psi) = (\partial {V'}/\partial t)$ then requires
\begin{equation}
\label{F, barotropic case}
F = \left[f(t) + g(\psi)\right]^{\frac{(a + b)}{a + b + (3n + 1)c}}.
\end{equation}
Substituting this expression into (\ref{Kasner-like metric in terms of F-bar}) and making the coordinate transformation $(df/dt)dt \rightarrow d\tilde{t}$, $(d g/d\psi)d\psi \rightarrow d\tilde{\psi}$, the line element in $5D$ can be written as
\begin{equation}
\label{5D line element for the barotropic case}
dS^2 = \frac{D d\tilde{t}^2}{\tilde{H}^{(3n + 1)c}} - A \tilde{H}^{2 a \alpha } dx^2 - B \tilde{H}^{2 a \beta } dy^2 - C \tilde{H}^{2 a \gamma } dz^2 - \frac{D  d\tilde{\psi}^2}{\tilde{H}^{(3n + 1)c}},
\end{equation}
where $\tilde{H} \equiv \left(\tilde{t} + E \tilde{\psi}\right)^{\frac{1}{a + b + (3n + 1)c}}$; $E$ is an arbitrary  constant for $n = 1/3$,  but $E = \pm 1$ for any other $n \neq 1/3$; $D$ is a positive constant introduced for dimensional considerations.

\subsubsection{Kasner universe in $5D$}

We immediately note that the case where   $E = 0$, which requires $n = 1/3$, gives back the well-known Kasner universe in $5D$. In fact, setting $\tilde{t} \propto \tau^{(a + b + 2c)/(a + b + c)}$  the line element (\ref{5D line element for the barotropic case}) reduces to 
\begin{equation}
\label{Kasner solution in 5D}
dS^2 = d\tau^2 - A \tau^{2q_{1}} dx^2 - B \tau^{2 q_{2}}dy^2 - C \tau^{2 q_{3}}dz^2 \pm D \tau^{2 q_{4}}d\psi^2, 
 \end{equation}
where
\begin{equation}
\label{defifition of the q's}
q_{1} = \frac{\alpha a}{(a + b + c)},  \;\;\;q_{2} = \frac{\beta a}{(a + b + c)}, \;\;\;q_{3} = \frac{\gamma a}{(a + b + c)}, \;\;\; q_{4} = - \frac{c}{a + b + c}, 
\end{equation}
 satisfy  $\Sigma_{i = 1}^4{q_{i}} = \Sigma_{i = 1}^4{q^2_{i}} = 1$, typical of  the Kasner universe in $5D$. In order to avoid misunderstandings, it may be useful to reiterate our terminology:   
 (i) the light-like Kasner metric is (\ref{Lightlike Kasner solution}), which depends on the light-like variable $u$; (ii) by   Kasner-like metric  we refer to (\ref{Kasner-like metric in terms of F-bar}), which depends on one arbitrary function of $t$ and $\psi$, and (iii) Kasner metric is the usual name given to  (\ref{Kasner solution in 5D}),  which depends only on $\tau$.  

\bigskip 

In general, for the $4$-dimensional interpretation of (\ref{5D line element for the barotropic case}) we should notice that on every hypersurface $\tilde{\psi} = \tilde{\psi}_{0} = $ constant (${\psi} = {\psi}_{0} = $ constant) the proper time $\tau$ is given by

\begin{equation}
\label{proper time}
d\tau = \pm \; \frac{\sqrt{D} d\tilde{t}}{(\tilde{t} + E \tilde{\psi}_{0})^m}, \;\;\;\;m \equiv \frac{(3n + 1)c}{2 \left[a + b + (3n + 1)c\right]}. 
\end{equation}
Bellow we consider several cases.

\subsubsection{Anisotropic Milne universe}

If $m = 0$, then\footnote{We exclude  $c = 0$ because it corresponds to empty space, i.e., $T_{\mu\nu} = 0$.} 

\begin{equation}
\label{Milne universe} 
 n = - \frac{1}{3}.
\end{equation}
In terms of the proper time $\tau = \sqrt{D}(\tilde{t} + E  \tilde{\psi}_{0}) $, the metric induced on $4$-dimensional hypersurfaces $\Sigma_{\psi}$  can be written as 
\begin{equation}
\label{anisotropic milne universe }
ds^2 = dS^2_{\Sigma_{|_{\psi}}} = d\tau^2 - \bar{A} \tau^{p_{1}}dx^2 - \bar{B} \tau^{p_{2}}dy^2 - \bar{C} \tau^{p_{3}}dz^2,  
\end{equation}
where $p_{i}$ are the parameters defined in (\ref{introduction of p1, p2 and p3}) and  $\bar{A}, \bar{B}$ and $\bar{C}$ are some new constants. In addition, $n_{i}$, the ratios of the anisotropic stresses to the energy density (\ref{effective stresses, radiation-like solution}) are given by 
\begin{equation}
\label{the n's for Milne universe}
n_{x} = \frac{(\alpha - \beta - \gamma)}{a}, \;\;\; n_{y} = \frac{(\beta - \alpha - \gamma)}{a}, \;\;\;n_{z} = \frac{(\gamma - \alpha - \beta)}{a}, \;\;\;8\pi G \rho^{(eff)} = \frac{a^2 c}{(a + b)^2 \tau^2}
\end{equation}
For $\alpha = \beta = \gamma $ we find $n_{x} = n_{y} = n_{x} = - 1/3$ and consequently we recover Milne's universe, as expected.

\subsubsection{Anisotropic  de Sitter universe}

If  $m = 1$, then
\begin{equation}
\label{m = 1}
n = - \frac{1}{3} - \frac{2(\alpha^2 + \beta^2 + \gamma^2)}{3(\alpha \beta + \alpha \gamma + \beta \gamma)}. 
\end{equation}
From (\ref{proper time}) we get $(\tilde{t} + E\tilde{\psi}_{0} ) \propto e^{\pm \tau/\sqrt{D}}$. Taking the negative sign,  the induced metric in $4D$ can be expressed as
\begin{equation}
\label{barotropic case, 4D metric}
ds^2 = dS^2_{\Sigma_{|_{\psi}}} = d\tau^2 - \bar{A}e^{p_{1}\tau/\sqrt{D}}dx^2 - \bar{B}e^{p_{2}\tau/\sqrt{D}}dy^2 - \bar{C}e^{p_{3}\tau/\sqrt{D}}dz^2. 
\end{equation}
For this metric we find
\begin{equation}
\label{matter quantities for the de Sitter-like solution}
n_{x} = - \frac{(\beta^2 + \gamma^2 + \beta \gamma)}{c}, \;\;\;n_{y} = - \frac{(\alpha^2 + \gamma^2 + \alpha \gamma)}{c}, \;\;\;n_{z} = - \frac{(\alpha^2 + \beta^2 + \alpha\beta)}{c}, \;\;\;8 \pi G \rho^{(eff)} = \frac{a^2 c}{(a + b)^2 D}.
\end{equation}

In the case of isotropic expansion $(\alpha = \beta = \gamma)$ these equations yield $n_{x} = n_{y} = n_{z} = n = - 1$ and (\ref{barotropic case, 4D metric}) reduces to the familiar de Sitter metric with cosmological constant $\Lambda_{(4)} = 3/D$. An interesting conclusion here is that  an anisotropic universe can enter a phase of exponential expansion (inflation), without satisfying the classical ``false-vacuum" equation $p = - \rho$ (see (\ref{m = 1})).  

\subsubsection{Anisotropic power-law FRW universe} 

For $m \neq 1$, from (\ref{proper time}) we obtain
\begin{equation}
\label{proper time, general case}
(\tilde{t} + E \tilde{\psi}_{0}) = \left[\frac{(1 - m)}{\sqrt{D}}(\tau - \tau_{0})\right]^{1/(1 - m)},
\end{equation}
where $\tau_{0}$ is a constant of integration. Thus, the induced metric in $4D$ becomes
\begin{equation}
\label{non-empty Kasner-like universe} 
ds^2 = dS^2_{\Sigma_{|_{\psi}}} = d\tau^2 - \bar{A}\tau^{\alpha \kappa}dx^2 - \bar{B}\tau^{\beta \kappa}dy^2 - \bar{C}\tau^{\gamma \kappa}dz^2,
\end{equation}
where we have set $\tau_{0} = 0$;  $\bar{A}, \bar{B}, \bar{C}$ are constants with the appropriate units,  and $\kappa$ is given by

\begin{equation}
\label{definition of kappa}
\kappa \equiv \frac{4 a}{2(a + b) + (3n + 1)c} = \frac{4(\alpha + \beta + \gamma)}{2(\alpha^2 + \beta^2 + \gamma^2) +(3n + 1)(\alpha \beta + \alpha \gamma + \beta \gamma)}.
\end{equation}
We note that the denominator of $\kappa$ is non-zero because by assumption here $m \neq 1$, see (\ref{m = 1}).         
  The effective matter quantities are
\begin{equation}
\label{matter quantities for the barotropic case}
\bar{p}^{(eff)} = n \rho^{(eff)}, \;\;\;8 \pi G \rho^{(eff)} = \frac{\kappa^2 c}{4 \tau^2}
\end{equation}
We also find
\begin{equation}
\label{the n's for the barotropic case}
n_{x} = \frac{2\alpha +(3n - 1)(\beta + \gamma)}{2 a}, \;\;\;n_{y} = \frac{2\beta +(3n - 1)(\alpha + \gamma)}{2 a}, \;\;\;n_{z} = \frac{2\gamma +(3n - 1)(\alpha + \beta)}{2 a}.
\end{equation}
We note that for  $c = 0$, the space is empty $(\rho^{(eff)} = 0)$, and the line element (\ref{non-empty Kasner-like universe}) yields the well-known Kasner solution in $4D$. Besides, for $n = 1$ and $n = 1/3$ the above expressions reduce to those obtained for perfect fluid and radiation-like matter discussed in  sections $3.1$ and $3.2$, respectively.

\subsubsection{Isotropic expansion: spatially flat FRW universe}

The above expressions evidence the fact that for anisotropic expansion  the effective EMT behaves like a perfect fluid \underline{only} for $n = 1$. In contrast,  isotropic expansion allows perfect fluid for any value of $n$. In this case the $5D$ metric (\ref{5D line element for the barotropic case}) can be written as (we omit the tilde over $t$ and $\psi$)
\begin{equation}
\label{barotropic case with isotropic expansion}
dS^2 = \frac{ D dt^2}{(t + E \psi)^{(3n + 1)/(3n + 2)}} - C (t + \psi)^{2/(2 + 3n)}\left[dx^2 + dy^ 2 + dz^2\right] - \frac{ D d\psi^2}{(t + E \psi)^{(3n + 1)/(3n + 2)}}.
\end{equation}
For $n \neq - 1$, on every hypersurface $\Sigma_{\psi}$ it reduces to 
\begin{equation}
\label{metric for the flat FRW model}
ds^2 = d\tau^2 - C \tau^{4/3(n + 1)}\left[dx^2 + dy^2 + dz^2\right],
\end{equation}
which is the familiar flat FRW model with perfect fluid 
\begin{equation}
\label{matter for FRW flat model}
p = n \rho, \;\;\; 8\pi G \rho= \frac{4}{3(n + 1)^2 \tau^2}. 
\end{equation}
For $n = - 1$ we recover the de Sitter spacetime as shown in (\ref{barotropic case, 4D metric}).

To finish this section we would like to emphasize that although the metrics with $a = 0$ and $c = 0$ correspond to empty space (Ricci-flat in $4D$), they are different in nature. For $a = 0$ the spacetime is Minkowski (Riemann-flat) in $5D$ and $4D$, while  for $c = 0$ the components of the Riemann tensor are nonzero  in $5D$ and in the  $4D$ subspace $\Sigma_{\psi}$.

\section{Cosmological models in $4D$. The braneworld approach}

The preceding discussion shows that, in the framework of STM  the  Kasner-like metric (\ref{Kasner-like metric in terms of F-bar})  embeds  a large family of  $4D$ cosmological  models that are  anisotropic versions of the FRW ones.    
However, one could argue that  the effective matter quantities (\ref{EMT}) do not have to satisfy the regular energy conditions \cite{Bronnikov}, or any physically motivated equation of state, because they involve terms of  geometric origin\footnote{In fact, the effective EMT defined  by (\ref{4D Einstein with T and K}) contains a contribution, given by    $E_{\alpha\beta}$, which  is the spacetime projection of the $5D$ Weyl tensor and connects the physics in $4D$ with the geometry in $5D$.}. 

In this section we will see that  the $5D$ metric (\ref{Kasner-like metric in terms of F-bar}) can be completely determined if one imposes an equation of state on the  matter in the brane. Although the concept is the same as in section $3$, the physics here is {\it different}. Namely, in this approach the spacetime is a singular hypersurface and, for the $5D$ Kasner metrics under consideration, there is an effective  non-vanishing cosmological term in $4D$ (the brane).  As a consequence,   the time evolution as well as the  interpretation of the solutions in $4D$ is distinct from the one obtained, under similar conditions,  in the framework of STM.

\subsection{The braneworld paradigm}

In order to make the paper self-consistent, and set the notation, we give a brief sketch of the technical details that we need in our discussion. In the  simplest RS2 braneworld scenario our universe is identified with a {\it fixed} singular hypersurface ${\Sigma_{\psi_{b}}}$ (called {\it brane}) embedded in a   $5$-dimensional  bulk  with ${\bf Z}_{2}$ symmetry  with respect to the brane. 
The discontinuity of the extrinsic curvature across $\Sigma_{\psi_{b}}$ is related to the presence of matter on the brane, which is  described by an EMT that we denote as $\tau_{\mu\nu}$. Thus, now the Einstein field equations in $5D$  are $G_{AB} = k_{(5)}^2 T_{AB}^{(brane)}$, where $k_{(5)}^2$ is a constant with the appropriate units and $T_{AB}^{(brane)} = \delta_{A}^{\mu}\delta_{B}^{\nu}\tau_{\mu\nu} \delta(\psi)/\Phi$.

Israel's boundary conditions \cite{Israel} relate the jump of $K_{\mu\nu}$ to $\tau_{\mu\nu}$, namely,
\begin{equation}
({K_{\mu \nu}}_{|\Sigma^{+}_{\psi_{b}}} - {K_{\mu \nu}}_{|\Sigma^{-}_{\psi_{b}}}) = - \epsilon \frac{k_{(5)}^2}{2}(\tau_{\mu\nu} - \frac{1}{3}\tau g_{\mu\nu}). 
\end{equation}
 Now, the assumed ${\bf{Z}}_{2}$ symmetry implies ${K_{\mu \nu}}_{|\Sigma^{+}_{\psi_{b}}}  = - {K_{\mu \nu}}_{|\Sigma^{-}_{\psi_{b}}}$. Consequently, 
 
\begin{equation}
\label{emt on the brane in terms of K}
\tau_{\mu\nu} = - \frac{2\epsilon}{k_{(5)}^2}\left(K_{\mu\nu} - g_{\mu\nu} K\right),
\end{equation}
where the extrinsic curvature $K_{\mu\nu}$ has to be evaluated on ${\Sigma^{+}_{\psi_{b}}}$.
From $G_{\nu 4} = 0$ it follows that $\tau^{\mu}_{\nu;\mu} = 0$.
Thus $\tau_{\mu\nu}$ represents the total, vacuum plus matter,  conserved energy-momentum tensor on the brane. It is usually separated in  two parts \cite{Csaki}, 
\begin{equation}
\label{decomposition of tau}
\tau_{\mu\nu} =  \sigma g_{\mu\nu} + {\cal{T}}_{\mu\nu},
\end{equation} 
where $\sigma$ is the tension of the brane, which is interpreted as the vacuum energy density, and ${\cal{T}}_{\mu\nu}$ represents the energy-momentum tensor of {\it ordinary} matter in $4D$. 

From (\ref{emt on the brane in terms of K}) and (\ref{decomposition of tau}) we get
\begin{equation}
\label{K in terms of matter in the brane}
K_{\mu\nu} = -  \frac{\epsilon k_{(5)}^2}{2} \left({\cal{T}}_{\mu\nu} - \frac{1}{3}g_{\mu\nu}({\cal{T}} + \sigma)\right).
\end{equation}
Substituting this expression into (\ref{4D Einstein with T and K}) we obtain \cite{Shiromizu}
\begin{equation}
\label{EMT in brane theory}
^{(4)}G_{\mu\nu} =  {\Lambda}_{(4)}g_{\mu\nu} + 8\pi G {\cal{T}}_{\mu\nu} - \epsilon k_{(5)}^4 \Pi_{\mu\nu} - \epsilon E_{\mu\nu},
\end{equation}
where
\begin{equation}
\label{definition of lambda}
\Lambda_{(4)} = - \epsilon \frac{k_{(5)}^4 \sigma^2}{12},
\end{equation}
\begin{equation}
\label{effective gravitational coupling}
8 \pi G =  - \epsilon \frac{k_{(5)}^4 \sigma}{6},
\end{equation}
and
\begin{equation}
\label{quadratic corrections}
\Pi_{\mu\nu} =  \frac{1}{4} {\cal{T}}_{\mu\alpha}{\cal{T}}^{\alpha}_{\nu} - \frac{1}{12}{\cal{T}} {\cal{T}}_{\mu\nu} - \frac{1}{8}g_{\mu\nu}{\cal{T}}_{\alpha\beta}{\cal{T}}^{\alpha\beta} + \frac{1}{24}g_{\mu\nu}{\cal{T}}^2.
\end{equation}
All these four-dimensional quantities have to be evaluated on ${\Sigma^{+}_{\psi_{b}}}$.  They contain two important features; they give a working definition of the fundamental quantities $\Lambda_{(4)}$ and $G$ and contain  higher-dimensional modifications to general relativity. Namely, local quadratic energy-momentum corrections via the tensor $\Pi_{\mu\nu}$, and the nonlocal effects from the free gravitational field in the bulk, transmitted  by $E_{\mu\nu}$.

\subsection{Matter in the brane. Gaussian coordinates}

In the braneworld literature the use of Gaussian coordinates in quite common. In these coordinates the function $F$ is given by (\ref{F in Gaussian coordinates, c2 = 0}), which under the re-scaling $F \rightarrow \bar{F}^{a^2/(a + b)}$ becomes $\bar{F} = l(t) + \psi h(t)$.  

If we locate the brane at $\psi = 0$, then the metric of the bulk is given by:

\begin{enumerate}
\item For $\psi > 0$
\begin{equation}
\label{bulk metric for psi positive}
dS^2_{(+)} = \frac{\left[\dot{l} + \psi \dot{h}\right]^2}{h^2}dt^2 - A \left[l(t) + \psi h(t)\right]^{p_{1}}dx^2 -  B \left[l(t) + \psi h(t)\right]^{p_{2}}dy^2 - C \left[l(t) + \psi h(t)\right]^{p_{3}}dz^2 - d\psi^2.
\end{equation}
\item For $\psi < 0$
\begin{equation}
\label{bulk metric for psi negative}
dS^2_{(-)} = \frac{\left[\dot{l} - \psi \dot{h}\right]^2}{h^2}dt^2 - A \left[l(t) - \psi h(t)\right]^{p_{1}}dx^2 -  B \left[l(t) - \psi h(t)\right]^{p_{2}}dy^2 - C \left[l(t) - \psi h(t)\right]^{p_{3}}dz^2 - d\psi^2.
\end{equation}

\end{enumerate}
Using (\ref{extrinsic curvature}) we calculate the non-vanishing components of $K_{\mu\nu} = {K_{\mu \nu}}_{|\Sigma^{+}_{\psi_{b}}}$. These are  
\begin{equation}
\label{extrinsic curvature in Gaussian coordinates}
K_{00} = \frac{\dot{l}\dot{h}}{h^2}, \;\;\;K_{11} = - \frac{A \alpha a l^{(p_{1} - 1)}h}{(a + b)},\;\;\;K_{22} = - \frac{B \beta a l^{(p_{2} - 1)}h}{(a + b)}, \;\;\;K_{33} = - \frac{C \gamma a l^{(p_{3} - 1)}h}{(a + b)}.
\end{equation}
We assume that the matter in the brane satisfies the equation of state
\begin{equation}
\label{equation of state for ordinary matter}
p = n \rho,
\end{equation}
where $\rho = {\cal{T}}_{0}^{0}$, $p = (p_{x} + p_{y} + p_{z})/3$ and $p_{x} = - {\cal{T}}_{1}^{1}$, $p_{y} = - {\cal{T}}_{2}^{2}$, $p_{z} = - {\cal{T}}_{3}^{3}$. Using these expressions, from (\ref{emt on the brane in terms of K}), with $\epsilon = - 1$,  and (\ref{decomposition of tau}) we obtain

\begin{eqnarray}
\label{ordinary matter quantities}
k_{(5)}^2 \sigma &=& - \frac{2}{(1 + n)}\left[\frac{\dot{h}}{\dot{l}} + \frac{(2 + 3n) a^2}{3(a + b)}\left(\frac{h}{l}\right)\right], \nonumber \\
k_{(5)}^2 \rho &=&  \frac{2}{(1 + n)}\left[\frac{\dot{h}}{\dot{l}} - \frac{a^2}{3 (a + b)}\left(\frac{h}{l}\right)\right], \;\;\;n \neq  -1, \;\;\;\dot{l} \neq 0.
\end{eqnarray}
We notice that in cosmological applications  the metric function $g_{00}$ is subjected to the condition \cite{Binetruy}, \cite{strength}
\begin{equation}
\label{condition on the metric}
{g_{00}}_{|_{brane}} = 1.
\end{equation}
Thus
\begin{equation}
\label{condition on h and l}
h(t) = s \; \dot{l}(t), \;\;\;s = \pm 1.
\end{equation}
Therefore,  we have two equations for the three unknown $\sigma, \rho$ and $l(t)$. Taking the covariant divergence of (\ref{decomposition of tau}), it follows that to conserve both the total brane energy-momentum tensor $\tau_{\mu\nu}$ and the matter energy-momentum tensor ${\cal{T}}_{\mu\nu}$, we must have $\sigma = \sigma_{0} = $ constant. Then,  using (\ref{condition on h and l})  we integrate the first equation in (\ref{ordinary matter quantities})  and obtain the scale factor as\footnote{From (\ref{effective gravitational coupling}), with $\epsilon = - 1$, it follows that  $\sigma$ must be positive in order to ensure $G > 0$.} 
  
\begin{equation}
\label{exponential expansion}
l(t) = \left[C_{1} e^{- s(n + 1)k_{(5)}^2 \sigma_{0} t/2} + C_{2}\right]^{\eta}, \;\;\; \eta \equiv \frac{3(a + b)}{(2 + 3n)a^2 + 3(a + b)}, 
\end{equation}
where $C_{1}$ and $C_{2}$ are constants of integration. We note that $\eta$ is positive for arbitrary values of $\alpha$, $\beta$, $\gamma$ and $n > -1$. Therefore, if  we choose $s = - 1$ and set   $C_{2} = 0$, then the ``origin" $l = 0$ is located at $t = - \infty$. Thus, from (\ref{ordinary matter quantities}) we find $\rho \propto \sigma = \sigma_{0}$ for all $t$, regardless of the value of $n$ (but  $n \neq - 1$). The resulting metric on the brane is de Sitter-like, with different rates of exponential expansion in every direction, similar to the models discussed in (\ref{barotropic case, 4D metric}).

\subsection{Non-Gaussian embeddings}

The question may arise of whether the simplicity of the above scenario is not a consequence of the simplifying assumption of Gaussian coordinates. 
In order to investigate this question, we consider here the  embedding that  arises from the choice\footnote{This is a  simple combination between $\dot{V} \propto \dot{F}F^{[c(1 + 3n)/(a + b)]}$ for the anisotropic FRW  models considered in Section $3.3$,  and $\dot{V} \propto \dot{F}/F'^2$ for the Gaussian embedding discussed in Section $4.2$.}

\begin{equation}
\label{generalization of V dot}
\dot{V} \propto \frac{\dot{F} F^{q c/(a + b)}}{F'^2},
\end{equation}
where $q$ is some constant.  With this choice the metric in $5D$ can be written as

\begin{equation}
\label{metric for sigma constant without exponential expansion}
dS^2 = \frac{ F^{q c/(a + b)} \dot{F}^2}{F'^2} dt^2 - A F^{p_{1}}dx^2 - B F^{p_{2}}dy^2 - C F^{p_{3}}dz^2 -  F^{q c/(a + b)} d\psi^2.
\end{equation}
Now, using $(\partial \dot{V}/\partial \psi) = (\partial {V'}/\partial t)$ we find
\begin{equation}
\label{F generalized}
F = \left[l(t) + \psi h(t)\right]^{(a + b)/(a + b - q c)}, \;\;\;\mbox{for}\;\;\; q \neq \frac{a + b}{c}, 
\end{equation}
and 
\begin{equation}
\label{F for Generalization II}
F = l(t)e^{\psi h(t)},\;\;\; \mbox{for}\;\;\; q = \frac{a + b}{c},
\end{equation}
where  $l(t)$ and $h(t)$ are arbitrary functions of integration. Again, if we locate the brane at $\psi = 0$, then the metric  in the ${\bf{Z}}_{2}$-symmetric bulk is obtained by replacing  $\psi \rightarrow |\psi|$ in (\ref{F generalized}) and (\ref{F for Generalization II}).
Following  the steps used in Section $(4.2)$ we find
\begin{equation}
\label{equation for sigma in the generalized model}
k_{(5)}^2 \sigma = - \frac{2 s }{n + 1}\left[\frac{\ddot{l}}{\dot{l}} + \frac{(2 + 3 n)a^2 + 3 q c}{3 (a + b - q c)}\;\left(\frac{\dot{l}}{l}\right)\right], \;\;\;\;q \neq \frac{a + b}{c}.
\end{equation}
It is interesting to note that in the case where $q = (a + b)/c$, the equation for $\sigma$ can {\it formally} be obtained from  (\ref{equation for sigma in the generalized model}) by  setting $q = 0$. Consequently, (\ref{F for Generalization II}) yields models on the brane that are identical to those in  Gaussian coordinates (\ref{ordinary matter quantities}), although the metric in the $5D$ bulk is completely different in both cases. 

The conclusion emanating from (\ref{equation for sigma in the generalized model}) is that, within the context of the $5$-dimensional Kasner spacetimes under consideration,  Gaussian and non-Gaussian  embeddings  generate the same  physics on the brane.    In particular, the assumption of constant $\sigma$, which is equivalent to a constant cosmological term $\Lambda_{(4)}$, obliges the universe to expand in a de Sitter {\it anisotropic} form regardless of (the choice of) the embedding. This is quite analogous to the cosmological ``no-hair" theorem/conjecture  of general relativity.

\subsection{Vacuum solutions on the brane}

Since the extra dimension is spacelike, the solutions to the field equations are invariant under the transformation $(x, y, z) \leftrightarrow \psi$. However, the physics in $4D$ crucially depends on how we choose our ordinary $3D$ space. 

In order to illustrate this, let us permute $\psi \leftrightarrow z$ in the solution given (\ref{metric for sigma constant without exponential expansion}) and (\ref{F generalized}). Also, to avoid misunderstanding we change $F(t, \psi) \rightarrow H(t, z)$. Using this notation, we find that  the metric

\begin{equation}
\label{vaccum solutions on the brane}
dS^2 = \frac{ H^{q c/(a + b)} \dot{H}^2}{H_{z}^2} dt^2 - A H^{p_{1}}dx^2 - B H^{p_{2}}dy^2 -  H^{q c/(a + b)} dz^2 \pm C H^{p_{3}}d\psi^2,
\end{equation}
where $H_{z} \equiv \partial H/\partial z$ and $H = \left[l(t) + z h(t)\right]^{(a + b)/(a + b - q c)}$, is also a   solution of the field equations $R_{AB} = 0$. Although (\ref{metric for sigma constant without exponential expansion}) and (\ref{vaccum solutions on the brane}) are diffeomorphic in $5D$, their interpretation in $4D$ is quite different. Specifically, unlike (\ref{metric for sigma constant without exponential expansion}) in (\ref{vaccum solutions on the brane}): (i) the extra dimension can be either spacelike or timelike, (ii) the spacetime slices $\Sigma_{\psi}$ are non-flat,  and (iii) the metric of the spacetime is independent of  $\psi$.  As a consequence of the  latter,   the extrinsic curvature $K_{\mu \nu}$, defined by (\ref{extrinsic curvature}), vanishes identically. Which in turn, by virtue of (\ref{emt on the brane in terms of K}), implies $\tau_{\mu\nu} = 0$, i.e., the spacetime (the brane) is devoid of matter $({\cal{T}}_{\mu\nu} = 0)$ and $\Lambda_{(4)} = 0$.

Clearly, other $5D$ metrics with properties  similar to (\ref{vaccum solutions on the brane}) can be constructed from the solutions (\ref{the metric in Gaussian coordinates})-(\ref{F in the synchronous coordinates}) of section $2$ as well as from (\ref{F for perfect fluid}) and (\ref{F, barotropic case}) of section $3$. 
The conclusion  here is that the spacetime part of  the   $5D$ Kasner-like metric (\ref{Kasner-like metric in terms of F-bar}),  after the  transformation    $\psi \leftrightarrow (x, y, z)$, can be interpreted as vacuum solutions in a braneworld ${\bf{Z}}_{2}-$symmetric scenario.

\section{Static embeddings }

As we noted above,  when a $5D$ metric is independent of $\psi$, the extra dimension can be either spacelike or timelike. This is a general feature of the $5D$ field equations \cite{XtraSym}. Therefore, after the transformation $\psi \leftrightarrow t$ the $5D$ metric still satisfies the field equations $R_{AB} = 0$.  The interesting feature here  is that after such a transformation the line element  induced on  $4D$ hypersurfaces $\Sigma_{\psi}$ is static, instead of dynamic as in sections $3$ and $4$. 

A simple $5D$ line element  that illustrates this feature, in a quite general way, can be obtained from  the Kasner-like  metric (\ref{Kasner-like metric in terms of F-bar}) in the synchronous reference system.  In fact, making the transformations ${\psi \leftrightarrow z}$,  $F(t, \psi) {\rightarrow} H(t, z)$;  and $\psi \leftrightarrow t$, $H(t, z) \rightarrow W(\psi, z)$,  from (\ref{cosmological homogeneous solution in synchronous coordinates }) and (\ref{F in the synchronous coordinates}) we obtain 
\begin{equation}
\label{static solution}
dS^2 =  C W^{p_{3}} dt^2 - A W^{p_{1}}dx^2 - B W^{p_{2}}dy^2 - \left(\frac{W_{z}}{W'}\right)^2 dz^2  + d\psi^2, 
\end{equation}
 where (We recall the re-scalling of $F$ introduced at the end of section $2$.)
\begin{equation}
\label{the function S}
W = M(z) + \psi N(z). 
\end{equation}
This is a solution of the $5D$ equations $R_{AB} = 0$ for any arbitrary functions   $M(z)$ and $N(z)$.  It  explicitly depends on the extra dimension $\psi$, which now is timelike.  At this point it is worthwhile to emphasize that in modern noncompactified $5D$ theories both, spacelike and timelike extra dimensions are  physically admissible \cite{epsilon}. 

Once again the choice of the functions $M(z)$ and $N(z)$ depends on the version of $5D$ relativity we use to evaluate the properties  of the matter content in  $4D$. Bellow we illustrate this by considering the induced matter approach, used in section $3$, and the braneworld paradigm used in section $4$. 

\subsection{Static solutions with planar symmetry in conventional $4D$ general relativity}

It is not difficult to show that the components of the effective EMT,   induced on  spacetime hypersurfaces $\Sigma_{\psi}:\psi = \psi_{0} =$ constant,  for the metric (\ref{static solution}) satisfy algebraic relations similar, but not identical\footnote{For example (\ref{rel between the stresses}) is now replaced by $- (\alpha + \gamma)T_{1}^{1} + (\beta + \gamma)T_{2}^{2} + (\beta - \alpha)T_{3}^{3} = 0$.},  to those in (\ref{rel between the stresses}) and (\ref{rel between T0 and the stresses}), which are independent on the specific choice of $M$ and $N$. We omit them here and present the case where  the effective matter quantities satisfy the barotropic linear equation of state (\ref{barotropic equation of state}). In such a case we find
\begin{equation}
\label{M for the general static solution}
M(z) = \bar{C} N(z)^{k} - \psi_{0}N(z), \;\;\;k \equiv - \frac{(\alpha^2 + \beta^2 + \gamma^2)[(3n + 1)(\alpha + \beta)] + 2 \gamma]}{(\alpha \beta + \alpha \gamma + \beta \gamma)[(3n + 1)(\alpha + \beta - \gamma) + 4 \gamma]},
\end{equation}
where $\bar{C}$ is a constant of integration. Thus, in the $5D$ solution (\ref{static solution}):
$W = \bar{C} N^k + (\psi - \psi_{0})N$, which implies that the metric induced in $4D$ is independent  of the choice of the hypersurface $\Sigma_{\psi}$. The effective energy density in $4D$ is given by 
\begin{equation}
\label{static rho effective}
8 \pi G \rho^{(eff)} = \frac{2 a^2 \gamma c \; N^{2(1 - k)}}{\bar{C}^2 [(3n + 1)(\alpha + \beta) + 2\gamma](a + b)^2 },
\end{equation}
and the stresses are
\begin{equation}
\label{the static stresses}
\frac{p_{x}^{(eff)}}{\rho^{(eff)}} = \frac{(3n - 1)\gamma - \alpha (3n + 1)}{2 \gamma}, \;\;\;\frac{p_{y}^{(eff)}}{\rho^{(eff)}} = \frac{(3n - 1)\gamma - \beta (3n + 1)}{2 \gamma}, \;\;\;\frac{p_{z}^{(eff)}}{\rho^{(eff)}} = \frac{(3n + 1)(\alpha + \beta) + 2\gamma}{2 \gamma}.
\end{equation}
We note that for $k =  1$ these  quantities are constants and $\rho^{(eff)} < 0$ for all values of $\alpha, \beta$ and $\gamma$. In what follows we assume  $k \neq 1$.

Since ${g_{33}}_{|_{\Sigma_{\psi}}} = - {\bar{C}}^2 k^2 N^{2(k - 2)}(dN/dz)^2$ we can make the coordinate transformation  
 $N^{(k - 2)} dN \rightarrow d\bar{z}$, i.e., $N \sim {\bar{z}}^{1/(k - 1)}$.  In terms of this new coordinate the static metric in $5D$ is generated by (henceforth we omit the bar over $z$)
\begin{equation}
\label{S generating static solutions }
W(t, \psi) = \bar{C}\left[z^{k/(k - 1)} + (\psi - \psi_{0}) z^{1/(k - 1)}\right]
\end{equation}
The matter quantities induced in $4D$ decrease as $1/z^2$. Therefore, 
the above equations represent  static ``pancake-like" distributions  where the matter is concentrated  near the plane $z = 0$, while  far from it $\rho \rightarrow 0$. 

Except for the singularity at $z = 0$, the matter distribution presents ``reasonable" physical properties. Indeed,   for every value of $n $,  the ``physical" conditions  $\rho^{(eff)} > 0$ and  $\rho \geq |p_{x, y, z}|$ are satisfied in a wide range of parameters $\alpha, \beta$ and $\gamma$. As an illustration, in the case of axial symmetry with respect to $z$, for $n = 0$ these conditions hold in the range $- 2\beta/3 < \gamma < - \beta/2$ ($\alpha = \beta > 0$) or   $-\beta/2 < \gamma < - 2\beta/3$ $(\alpha = \beta < 0)$. For $n = 1/3$,  they hold  if $- 2 \beta < \gamma \leq - \beta$ ($\alpha = \beta > 0$) or  $- \beta \leq \gamma < - 2\beta$ ($\alpha = \beta < 0$). A similar analysis can be extended for other values of $n$.

A simpler solution can be obtained  from the above expressions in the limiting case where $k = \infty$, which occurs for $(3n + 1)(\alpha + \beta - \gamma) + 4\gamma = 0$. In this case (\ref{S generating static solutions }) simplifies to $W = \bar{C}\left[z + (\psi - \psi_{0})\right]$ and the matter quantities are obtained from (\ref{static rho effective}), (\ref{the static stresses}) by replacing $\gamma \rightarrow [(3n + 1)(\alpha + \beta)/3(n - 1)]$. In the case of axial symmetry, the line element becomes independent of the parameters $\alpha, \beta, \gamma$ and depends only on $n$. The effective matter quantities   satisfy  $\rho^{(eff)} > 0$ and  $\rho \geq |p_{x, y, z}|$ for any $n$ in the range $- 1/3 \leq  n <  1/3 $.

\subsection{Static solutions on the brane}

We now proceed to use the braneworld technique for evaluating the matter quantities. If we locate the brane at $\psi = 0$, then the metric in the ${{\bf{Z}}_{2}}$-symmetric bulk is generated by $W = M(z) + |\psi| N(z)$. From (\ref{emt on the brane in terms of K}), with $\epsilon =  1$,  and (\ref{decomposition of tau}) we obtain  the components of ${\cal{T}}_{\mu\nu}$. Now the barotropic equation of state 
(\ref{equation of state for ordinary matter})  yields  a differential equation linking $M(z), N(z)$ and $\sigma$, which can be easily integrated for constant vacuum energy, $\sigma = \sigma_{0}$. Namely, we obtain

\begin{equation}
\label{Static solutions on the brane, N(z)}
N(z) =  \frac{3 k_{(5)}^2\sigma_{0} (a + b)(1 + n) M(z)}{2(5 + 3n)[\gamma^2 + \gamma(\alpha + \beta)] + 4(2 + 3n)[\alpha^2 + \beta(\alpha + \beta)]} - E M(z)^{- \frac{a [(2 + 3n)(\alpha + \beta) + 3\gamma]}{(a + b)(2 + 3n)}},
\end{equation}
where $E$ is a constant of integration. Using this expression we obtain 
\begin{equation}
\label{Static solutions on the brane, rho}
k_{(5)}^2 \rho = \frac{6 E a \gamma}{(a + b)(2 + 3n) M(z)^{\tilde{k}}} + \frac{2 k_{(5)}^2 \sigma_{0}[\alpha^2 + \beta^2 - \gamma^2 + \alpha \beta
 - \gamma(\alpha + \beta)]}{\tilde{k}(a + b)(2 + 3n)},
\end{equation}
where
\begin{equation}
\label{Static solutions on the brane, definition of k tilde}
\tilde{k} = \frac{(5 + 3n)[\gamma^2 + \gamma(\alpha + \beta)] + 2(2 + 3n)[\alpha^2 + \beta(\alpha + \beta)]}{(\alpha^2 + \beta^2 + \gamma^2)(2 + 3n)}.
\end{equation}
We note that $\rho = $ constant for $\gamma = 0$. Therefore, in what follows we assume $\gamma > 0$.  Since $M(z)$ is an arbitrary function, without loss of generality we can choose it as\footnote{ We note that in the case under consideration  $(\gamma \neq 0)$, $\tilde{k}$ never vanishes. In fact, for real parameters $\alpha$ and $\beta$ the quantity $[\alpha^2 + \beta(\alpha + \beta)]$ is always positive. On the other hand, $[\gamma^2 + \gamma(\alpha + \beta)] = 0$ requires $\alpha + \beta + \gamma = 0$, i.e., $a = 0$, which corresponds to Minkowski space in $5D$ and $4D$.}  
\begin{equation}
\label{choice of M(z)}
M(z) \sim  z^{2/\tilde{k}}, 
\end{equation}
which is suggested by the decrease of the effective density discussed in section $5.1$. 
It is not difficult to see that $\rho$ is positive for a large number of values of $\alpha, \beta, \gamma$. The positivity of the first term is guaranteed by the constant of integration $E$.   To illustrate the positivity of the second term, we once again consider the case with axial symmetry with respect to $z$. In this case we find
\begin{equation}
\label{Static solutions on the brane, rho for axial symmetry}
\lim_{z \rightarrow  \pm \infty}{\rho} = \frac{2\sigma_{0} (3\beta + \gamma)(\beta - \gamma)} {\gamma (5 + 3n)(2\beta + \gamma) + 6\beta^2(3n + 2)},
\end{equation}
which is positive for any $\beta > \gamma$ and $n \geq - 2/3$.

The main conclusion from this section  is  that regardless of whether we use the braneworld paradigm or the induced matter approach, the basic picture in $4D$ is essentially the same. Namely that the $4D$ part of (\ref{static solution}) represents static pancake-like distributions of matter.

\section{Summary}

The vacuum Einstein field equations for the $5D$ FRW line element (\ref{Usual cosmological metric in 5D}) allow complete integration in  a number of cases. In particular for  $\Phi = 1$, or $n = 1$,  the $(t, \psi)$-component of the field equations provides a relation that leads to a set of first integrals \cite{Binetruy}, \cite{JPdeL-isotropicCosm}. However, for the simplest anisotropic extension of (\ref{Usual cosmological metric in 5D}), namely, the diagonal Bianchi type-I metric 
\begin{equation}
\label{Bianchi type I, conclusions}
dS^2 = n^2(t, \psi)dt^2 - \sum_{i = 1}^{3}b_{i}(t, \psi) \left(dx^{i}\right)^2 + \epsilon \Phi^2(t, \psi)d\psi^2,
\end{equation}
this procedure  does not work. (For a discussion, and a new point of view in the context of braneworld, see \cite{Antonio}.) 

Here we have pointed out that making the coordinate transformation $du \propto \left(n dt - \Phi d\psi\right)$, $dv \propto \left(n dt + \Phi d\psi\right)$,  in (\ref{Bianchi type I, conclusions}) with $\epsilon = - 1$, the field equations allow complete integration in several physical situations, viz., (\ref{new solution}), (\ref{solution 2}), (\ref{solution 3}). The $4D$ interpretation of the $5D$ solutions requires the introduction of coordinates adapted to spacetime sections $\Sigma_{\psi}$. We introduced the $(t, \psi)$ coordinates by setting $u = F(t, \psi)$ and $v = V(t, \psi)$, and used a foliation of the $5D$ manifold such that $\Sigma_{\psi}$ is a hypersurface of the foliation that  is  orthogonal to the extra dimension with tangent ${\hat{n}}^{A} = \delta^{A}_{4}/\Phi$.  
From a mathematical point of view the functions $F$ and $V$ can be arbitrary, except for the fact that they have to satisfy (\ref{diagonal 5D metric}). However,  from a physical point of view, they are related to two important aspects of the construction of the spacetime, namely: (i) the choice of coordinates in $5D$, e.g., Gaussian normal coordinates adapted to $\Sigma_{\psi}$,  and (ii)  the formulation of physical conditions on the matter fields in $4D$, e.g., some an equation of state. 

Our study shows that there is a great freedom for embedding a $4D$ spacetime in an anisotropic $5D$ cosmological model. Similar results but in a distinct context have been found in \cite{Antonio}. To simplify the algebraic expressions, but not the physics, in sections $3$, $4$ and $5$ we have devoted our attention to the study of $4D$ spacetimes embedded in the light-like Kasner cosmological metric, which is a simplified version of (\ref{light-like metric in t-psi coordinates}).  

 We have seen that the simple one-variable line element (\ref{Lightlike Kasner solution}) can accommodate  a great variety of models in $4D$.  Indeed, within the context of STM and braneworld theories, we have shown here that  the Kasner-like metrics  (\ref{Kasner-like metric in terms of F-bar}) may be used or interpreted as embeddings for a large number of cosmological and static spacetimes in $4D$.  Thus, apparently ``different" astrophysical and cosmological scenarios in $4D$ might just be distinct versions of  the same physics in $5D$ \cite{XtraSym}.  

This investigation can be extended, or generalized, in different ways. In particular, we have not fully examined the possible $4D$ interpretation of the self-similar homothetic solution (\ref{new solution}). Neither, have we investigated the solutions (\ref{solution 2}) and (\ref{solution 3}). An important future development here is the question of how these solutions can be applied in the generalizations of Mixmaster or Belinskii-Khalatnikov-Lifschits oscillations, as well as other issues mentioned in section $1$, which appear in theories with one extra dimension.

\paragraph{Acknowledgments:} I wish  to thank one of the anonymous referees for a careful reading of this manuscript as well as for helpful and constructive suggestions.

\renewcommand{\theequation}{A-\arabic{equation}}
  \setcounter{equation}{0}  
  \section*{Appendix: Solving the field equations. Part II}  

Here we show two more families of analytic solutions to the field equations (\ref{Rxx, Ryy, Rzz}). With this aim we notice that  
these equations are greatly simplified if we introduce the function
\begin{equation}
\label{definition of calV}
{\cal{V}} = e^{\lambda + \mu + \sigma},
\end{equation}
in terms of which (\ref{Rxx, Ryy, Rzz}) become
\begin{eqnarray}
\label{Rxx, Ryy, Rzz in terms of calV}
4 \lambda_{u v} + \frac{\lambda_{u} {\cal{V}}_{v}}{{\cal{V}}} + \frac{\lambda_{v} {\cal{V}}_{u}}{{\cal{V}}} &=& 0, \nonumber \\
4 \mu_{u v} +  \frac{\mu_{u} {\cal{V}}_{v}}{{\cal{V}}} + \frac{\mu_{v} {\cal{V}}_{u}}{{\cal{V}}}&=& 0, \nonumber \\
4 \sigma_{u v} + \frac{\sigma_{u} {\cal{V}}_{v}}{{\cal{V}}} + \frac{\sigma_{v} {\cal{V}}_{u}}{{\cal{V}}} &=& 0.
\end{eqnarray}
Adding these equations and using (\ref{definition of calV}) we obtain an equation for ${\cal{V}}$, namely,
\begin{equation}
\label{equation for calV}
2 {\cal{V}}{\cal{V}}_{u v} - {\cal{V}}_{u} {\cal{V}}_{v} = 0,
\end{equation}
whose general solution can be written as

\begin{equation}
\label{solution for calV}
{\cal{V}} = \left[\tilde{h}(u) + \tilde{g}(v)\right]^2,
\end{equation}
where   $\tilde{h}$ and $\tilde{g}$ are arbitrary functions of their arguments. Clearly, the self-similar solution discussed in section $2.1$ corresponds to the particular choice 
\begin{equation}
\label{the metric functions in terms of calV}
e^{\lambda} \propto {\cal{V}}^{\alpha/a}, \;\;\;e^{\mu} \propto {\cal{V}}^{\beta/ a}, \;\;\;e^{\sigma} \propto {\cal{V}}^{\gamma/ a},
\end{equation} 
 which satisfies (\ref{Rxx, Ryy, Rzz in terms of calV}) and (\ref{equation for calV}) identically. 

In what follows, as in section $2.1$, we introduce a new set of null coordinates $\tilde{u}$ and $\tilde{v}$ by the relations  $\tilde{h}(u) = {\tilde{c}}_{1} \tilde{u}$ and $\tilde{g}(v) = {\tilde{c}}_{2} \tilde{v}$, where ${\tilde{c}}_{1}$ and ${\tilde{c}}_{2}$ are constants. In terms of these new coordinates ${\cal{V}} = \left({\tilde{c}}_{1} \tilde{u} + {\tilde{c}}_{2} \tilde{v}\right)^2$. Substituting this expression into the first of the equations (\ref{Rxx, Ryy, Rzz in terms of calV}), and dropping the tilde characters,  we obtain 
\[
2 \left(c_{1} u + c_{2} v\right)\lambda_{u v} + c_{1} \lambda_{v} + c_{2} \lambda_{u} = 0.
\]
A similar expression holds for $\mu$. The solutions bellow are obtained under the assumption that $e^{\lambda}$ and $e^{\mu}$ are separable functions of their arguments. In which case, the above equation implies that  they are proportional to $e^{\pm \left(c_{1} u - c_{2} v\right)}$. The metric function $e^{\sigma} = {\cal{V}}\; e^{- \left(\lambda + \mu\right)}$ automatically satisfies the third equation in (\ref{Rxx, Ryy, Rzz in terms of calV}) and is non-separable. 
Consequently, there are two different families of solutions corresponding to whether $e^{\lambda} \propto e^{-  \mu}$ or $e^{\lambda} \propto e^{  \mu}$. 

$\bullet$ In the case where $e^{\lambda} \propto e^{-   \mu}$, the field equations $R_{uu} = 0$ and $R_{vv} = 0$ reduce to
\begin{equation}
\label{separable 1}
  2 {\cal{A}}_{u}  - c_{1}\left(c_{1} u + c_{2} v\right){\cal{A}}  = 0,\;\;\;\mbox{and}\;\;\;  2 {\cal{A}}_{v}            - c_{2}\left(c_{1} u + c_{2} v\right){\cal{A}}  = 0. 
\end{equation}
These equations completely determine the function ${\cal{A}}$ and assure the fulfillment of  $R_{u v} = 0$. The final form of the solution is given by

\begin{equation}
\label{solution 2}
dS^2 = C_{0} e^{[\left(c_{1} u + c_{2}v\right)^2/4]} d u d v - C_{1} e^{\left(c_{1} u - c_{2} v\right)} d x^2 - C_{2} e^{- \left(c_{1} u - c_{2} v\right)} d y^2 
- \left(C_{1} C_{2}\right)^{- 1}\left(c_{1} u + c_{2} v\right)^2 d z^2.
\end{equation}

$\bullet$ Following the same steps as above we find that when $e^{\lambda} \propto e^{\mu}$ the solution is 
\begin{equation}
\label{solution 3}
dS^2 = C_{0} e^{[3 \left(c_{1} u + c_{2}v\right)^2/4]} e^{- 2\left(c_{1} u - c_{2} v\right)}d u d v - 
C \; e^{\left(c_{1} u - c_{2} v\right)} \left(d x^2 + d y^2\right) 
- C^{- 2} \left(c_{1} u + c_{2} v\right)^2 e^{- 2\left(c_{1} u - c_{2} v\right)} d z^2.
\end{equation}
In the above line elements  $\left(C, C_{0}, C_{1}, C_{2}\right)$ are constants of integration. We note that the resulting solutions are quite complicated even in the case where either $c_{1}$ or $c_{2}$  are set equal to zero. Although this is a great obstacle for the analytical interpretation of these metrics in $4D$, it allows us to appreciate the simplicity of the self-similar solutions discussed in the main text.

\end{document}